\documentclass{PoS}

\usepackage{amsmath,amssymb}
\usepackage{graphicx}
\usepackage{xspace}
\usepackage[small]{subfigure}
\usepackage{color}
\usepackage{multirow}

\newcommand{\nn}{\nonumber}

\newcommand{\be}{\begin{equation}}
\newcommand{\ee}{\end{equation}}

\newcommand{\vect}[1]{\mathbf{#1}}

\newcommand{\Lqcd}{\Lambda_{\text{QCD}}}

\newcommand{\GeV}{\text{ GeV}}

\newcommand{\as}{\alpha_s}

\newcommand{\cO}{\mathcal{O}}

\newcommand{\cI}{\mathcal{I}}

\newcommand{\cB}{\mathcal{B}}

\newcommand{\eq}[1]{Eq.~\eqref{#1}}

\newcommand{\fig}[1]{Fig.~\ref{fig:#1}}

\newcommand{\m}{{a}}
\newcommand{\B}{{b}}
\newcommand{\CM}{{c}}
\newcommand{\taun}{\tau_1}
\newcommand{\taum}{\tau_{1}^\m}
\newcommand{\tauB}{\tau_{1}^\B}
\newcommand{\tauCM}{\tau_{1}^\CM}

\newcommand{\taumBCM}{\tau_{1}^{\m,\B,\CM}}

\newcommand{\qBm}{q_B^\m}
\newcommand{\qJm}{q_J^\m}
\newcommand{\qBB}{q_B^\B}
\newcommand{\qJB}{q_J^\B}
\newcommand{\qBCM}{q_B^\CM}
\newcommand{\qJCM}{q_J^\CM}

\newcommand{\cumulant}{{\rm c}}
\newcommand{\sigmac}{\sigma_\cumulant}

\title{
 1-Jettiness in DIS: Measuring 2 Jets in 3 Ways 
\thanks{LA-UR-13-26573,\, MIT-CTP 4488}}

\ShortTitle{1-Jettiness in DIS}

\author{\speaker{Daekyoung Kang}\\
        Center for Theoretical Physics,  Massachusetts Institute of Technology, Cambridge, MA 02139, USA\\
        E-mail: \email{kang1@mit.edu}}

\author{Christopher Lee\\
        Theoretical Division, MS B283, Los Alamos National Laboratory, Los Alamos, NM 87545, USA\\
        E-mail: \email{clee@lanl.gov}}

\author{Iain W. Stewart\\
        Center for Theoretical Physics,  Massachusetts Institute of Technology, Cambridge, MA  02139, USA\\
        E-mail: \email{iains@mit.edu}}

\abstract{We compute cross sections for two-jet production in deep
  inelastic scattering (DIS), with one jet from initial state radiation (ISR)
  and the other from final state radiation, with a summation of large
  logarithms up to next-to-next-to-leading logarithmic (NNLL) accuracy.
  Use of the DIS event shape 1-jettiness ensures that events have two
  well-collimated jets. We calculate distributions for three versions of 1-jettiness that
 have different sensitivity to the transverse momentum of the ISR, and derive 
 factorization theorems for each of them using the soft collinear effective theory
  (SCET).  The structure of the transverse momentum dependence in the 
  factorization theorems is different for each 1-jettiness.  We
  present numerical results for these three observables with parameters
  for the HERA collider.  }

\FullConference{XXI International Workshop on Deep-Inelastic Scattering and Related Subject -DIS2013,\\
		22-26 April 2013\\
		Marseilles,France}

\begin{document}

 Studying jet production with event shapes can be advantageous, since
 it is possible to achieve higher precision compared to exclusive jet
 cross sections defined with jet algorithms.  In $e^+e^-$ collisions a
 classical example is thrust $T=1-\tau$~\cite{Farhi:1977sg}, where the
 hadronic final state is constrained to two jets for small $\tau$.
 Here results are available up to
 N$^3$LL$+\cO(\as^3)$~\cite{GehrmannDeRidder:2007bj,GehrmannDeRidder:2007hr,Weinzierl:2008iv,Weinzierl:2009ms,Becher:2008cf,Chien:2010kc,Abbate:2010xh,Abbate:2012jh}
 and this allows $\sim 1\%$ level theoretical accuracy in $\as$
 extractions.  A version of DIS thrust has been studied in HERA
 experiments
 \cite{Adloff:1997gq,Adloff:1999gn,Aktas:2005tz,Breitweg:1997ug,Chekanov:2002xk,Chekanov:2006hv}
 and was calculated in \cite{Antonelli:1999kx,Dasgupta:2002dc} at
 NLL$+\cO(\as^2)$ accuracy.  However, the HERA definition of the DIS thrust
 introduces a technical obstacle in theoretical calculations beyond
 NLL accuracy because it leads to non-global logarithms and it is
 unknown how to resum these beyond the leading
 logs~\cite{Dasgupta:2002dc,Dasgupta:2001sh}.  To determine higher
 order results for the log resummation and to rigorously include power
 corrections it is useful to derive factorization theorems that
 account for results to all orders in perturbation theory as well as
 the leading power corrections.  To do this we use the event shape
 1-jettiness, which is a thrust-like event shape without non-global
 logarithms.  We define three versions of 1-jettiness in Sec.~1,
 review results for the factorization theorems for these observables
 in Sec.~2, and give numerical results for 1-jettiness distributions
 up to NNLL accuracy in Sec.~3~\cite{Kang:2013nha}.


\section{$1$-Jettiness for DIS}
The 1-jettiness is a special case of $N$-jettiness introduced in
Ref.~\cite{Stewart:2010tn}.  The $N$-jettiness is a generalization of
thrust and a small value of $N$-jettiness constrains the final state to contain
$N+N_B$ jets where $N_B$ is the number of initial state jets
by ISR from proton beams and $N$ is the number of final state jets. In
DIS $N_B$ is 1.  In this paper we will focus on the case of a single final
jet ($N=1$), for which the DIS 1-jettiness observable is defined by
\be
\label{tau1intro}
\tau_1 = \frac{2}{Q^2}\sum_{i\in X} \min \{ q_B \cdot p_i, q_J\cdot p_i\}\,.
\ee
Here a four vector $q_B$ points along the incident proton momentum
and a four vector $q_J$ is picked to determine an axis for the measurement of the final-state jet.  
The min chooses the smaller scalar product, and also groups all particles in the final state $X$ into two regions,
particles closer to $q_B$ (smaller $q_B\cdot p_i$) and those closer to
$q_J$ (smaller $q_J\cdot p_i$).  This grouping depends on the choice of
$q_{B}$ and $q_J$. We consider three different cases:
\begin{subequations}
\label{tauABC}
\begin{align}
\taum: \qquad \qBm &= xP \,, & \qJm &= \text{jet axis} \\ \tauB:
\qquad \qBB &= xP \,, & \qJB &= q+xP \\ \tauCM: \qquad \qBCM &=P \,, &
\qJCM &= k\,,
\end{align}
\end{subequations}
where $P$, $k$, and $q$ are the initial proton, incoming electron, and
virtual boson momenta, respectively, and $x=Q^2/(2P\cdot q)$ is the
Bj\"{o}rken scaling variable where $q^2=-Q^2$.  The three variants
$\taumBCM$ in \eq{tauABC} are named for the corresponding properties
of the vector $q_{J}$.  In $\taum$, $\qJm$ is \emph{aligned} with a
jet axis that is defined either by a jet algorithm, or by a
minimization of the 1-jettiness in \eq{tau1intro} as
in~\cite{Thaler:2011gf}.  In $\tauB$, the vectors $\qJB$ and $\qBB$
are back-to-back in the \emph{Breit frame}.  
$\tauB$ can be rewritten 
in a way that is equivalent to one of the measured DIS thrust distributions except for the
normalization~\cite{Antonelli:1999kx,Kang:2013nha} and it could
 be analyzed with existing thrust data from the HERA experiment.
Similarly, for $\tauCM$ the vectors $q_B^c$ and $q_J^c$ are
back-to-back in the \emph{center-of-momentum frame}.

\section{Factorization Theorems for Different Jet Axes}

All orders factorization theorems for the three versions of
1-jettiness in \eq{tauABC} can be used to obtain higher order resutls,
and here we briefly describe them and highlight their differences. A
complete derivation and further details can be found in
\cite{Kang:2013nha}.  A factorization theorem for $\taum$ also has been
obtained in \cite{Kang:2012zr,Kang:2013wca}.  The cross section for
the three cases can be obtained as special cases of the general result
\begin{align} \label{factorizationintro}
\frac{d\sigma}{dx\,dQ^2\,d\tau_1}
& = \frac{d\sigma_0}{dx\,dQ^2} \, \int dt_J\, dt_B\, dk_S\, d^2\vect{p}_\perp 
\: \delta\Bigl( \tau_1 - \frac{t_J}{s_J} 
   - \frac{t_B}{s_B} - \frac{k_S}{Q_R}\Bigr)
   \nn \\
&\quad\times
\sum_{\kappa}  H_\kappa(Q^2,\mu)\,\,
 J_q(t_J - (\vect{q}_\perp + \vect{p}_\perp)^2,\mu)\,\,
 \cB_{\kappa/p}(t_B,x,\vect{p}_\perp^2,\mu) \,\,
  S_{\text{hemi}}(k_S,\mu) 
  \,,
\end{align}
where $\sigma_0$ is the Born cross section, $s_J,s_B,Q_R$ are
normalization constants which are different for $\taumBCM$ (see
\cite{Kang:2013nha}), and $\kappa$ is quark/antiquark flavors.
$H_\kappa$ is a hard function containing virtual corrections, and
determined by matching QCD onto SCET. $J_q$ is a quark jet function
describing radiation of collinear quarks and gluons from an initial
quark.  ${\cal B}_{\kappa/p}$ is a quark beam
function~\cite{Stewart:2009yx,Mantry:2009qz,Jain:2011iu} with a perturbative kernel
for collinear radiation and the parton distribution function (PDF) as
\begin{align} 
\cB_{\kappa/p}(t_B,x, \vect{p}_\perp^2,\mu)
  =\! \sum_j\!\!\int_x^1\! \frac{dz}{z}\:
\cI_{\kappa j}\Big(t_B,\frac{x}{z},\vect{p}_\perp^2,\mu\Big)\, f_{j/p}(z,\mu),  
\end{align}
where $t_B$ is the transverse virtuality ($p^+p^-$) of the quark $\kappa$,
and $\vect{p}_\perp$ is a transverse momentum of initial state radiation (ISR).
$S_{\text{hemi}}$ is the hemisphere soft function that describes radiation of 
soft particles from initial and final states. Note that $S_{\text{hemi}}$
for the three observables $\taumBCM$ is the same, which can be proved by using 
rescaling invariance of soft Wilson lines. Finally,
$\vect{q}_\perp$ is the transverse momentum of the virtual boson respect
to the jet and beam axes $q_{B}$ and $q_J$ in \eq{tau1intro}.

\eq{factorizationintro} has different transverse momentum dependencies
for the three 1-jettinesses. In the case of $\taum$, $q_J$ is aligned
with the jet axis and the argument of the jet function
$t_J-(\vect{q}_\perp+\vect{p}_\perp)^2 \to t_J$.  Here the transverse
integral acts on the beam function and it becomes the ordinary beam
function defined in Ref.~\cite{Stewart:2009yx}.
For $\tauB$, $\vect{q}_\perp$
is zero because $q$ is written as a linear combination of $q_{B,J}$,
but both $J_q$ and $\cB_{\kappa/p}$ involve $\vect{p}_\perp$.  For
$\tauCM$ there is no simplification from
\eq{factorizationintro}. Because of the different transverse momentum
dependence in the beam function for $\tauB$ and $\tauCM$, the
difference between these observables is sensitive to the transverse
momentum of the ISR.

Hard, jet, beam, and soft functions in \eq{factorizationintro} each
depend on a factorization scale $\mu$ (which is also precisely the
renormalization scale in SCET).  These functions contains logs of
$\mu^2/Q^2$, $\mu^2/(\tau_1 Q^2)$, or $\mu^2/(\tau_1^2 Q^2)$ and there
are always large logs when $\tau_1\ll 1$.  The large logs should be
resummed to achieve accurate prediction and this is achieved by using
renormalization group evolution (RGE) in SCET.  Each function is
evolved from the natural scale $\mu_{H,J,B,S}$ where its logs are
minimized to a common (arbitrary) scale $\mu$.  This sums up towers of
logs of $\taun$ to all order in $\as$.  The logarithmic accuracy of
the resummation is determined by the $\as$ order of the anomalous
dimensions. In this paper, our result of 1-jettiness is given to NNLL
accuracy which requires 3-loop cusp anomalous dimension, 2-loop
anomalous dimensions, and complete 1-loop results for the hard, jet,
beam, and soft functions.

Nonperturbative effects in the soft function from gluons with momenta
$\sim \Lambda_{\rm QCD}$ can be accounted for via a shape function.
To illustrate how the nonperturbative effect deforms perturbative
result, we adopt a simple model function for the peak region ($\tau_1\sim 2\Lambda_\text{QCD}/Q$). In the
tail region ($2 \Lambda_\text{QCD}/Q \ll \tau_1 \ll 1$) the universality of the leading nonperturbative
corrections has been shown for various $e^+e^-$ event shapes and
collision
energies~\cite{Salam:2001bd,Lee:2006fn,Lee:2006nr,Mateu:2012nk} (for
earlier work
see~\cite{Dokshitzer:1995zt,Akhoury:1995sp,Korchemsky:1994is}).  This
universality is also valid for power corrections for the three results
considered in \eq{factorizationintro}.  In the tail region, the
dominant power corrections are determined by a single parameter
$\Omega_1^{a,b,c}$ as
\begin{align}\label{powercorrection}
\frac{d\sigma}{d\taun}
&=\frac{d\sigma^\text{pert}}{d\taun}-\frac{2\Omega_1}{Q_R} \frac{d^2\sigma^\text{pert}}{d\taun^2}+\cdots
\,,\end{align}
where we leave $x$ and $Q$ dependencies in the cross section implicit.
Note that $\Omega_1$ is rigorously defined as a matrix element of a product of
soft Wilson lines. In Ref.~\cite{Kang:2013nha}, we proved the
universality of $\Omega_1$ for the three 1-jettinesses in the presence of hadron mass effects:
\begin{align}\label{universality}
\Omega_1&=\Omega_1^\text{a}=\Omega_1^\text{b}=\Omega_1^\text{c}
\,.\end{align}
This prediction can be tested experimentally.

\section{Numerical results at NNLL}
\label{sec:results}

Lets consider numerical results for the three
1-jettiness: $\taum$, $\tauB$, and $\tauCM$.  The results are
accurate for small $\taun$, and are resummed to LL, NLL, or NNLL accuracy, and also include the
singular terms at fixed order $\cO(\as)$.  We present
the $\taum$ spectra first, and then compare $\tauB$ and
$\tauCM$ spectra to $\taum$.  For the total invariant mass the value $s=(300~\text{GeV})^2$
 in the H1 and ZEUS experiments is used.
We also present cumulant cross sections $\sigmac(\taun)$ which are 
defined as
\be
\label{cumulant}
\sigma_c(\taun,x,Q^2)
=\frac{1}{\sigma_0}\int_0^{\taun} d\taun' \frac{d\sigma}{dx\, dQ^2\, d\taun'}
\,,\ee
where $\sigma_0$ is the Born cross section.

\begin{figure}[t!]{
      \includegraphics[width=0.45\columnwidth]{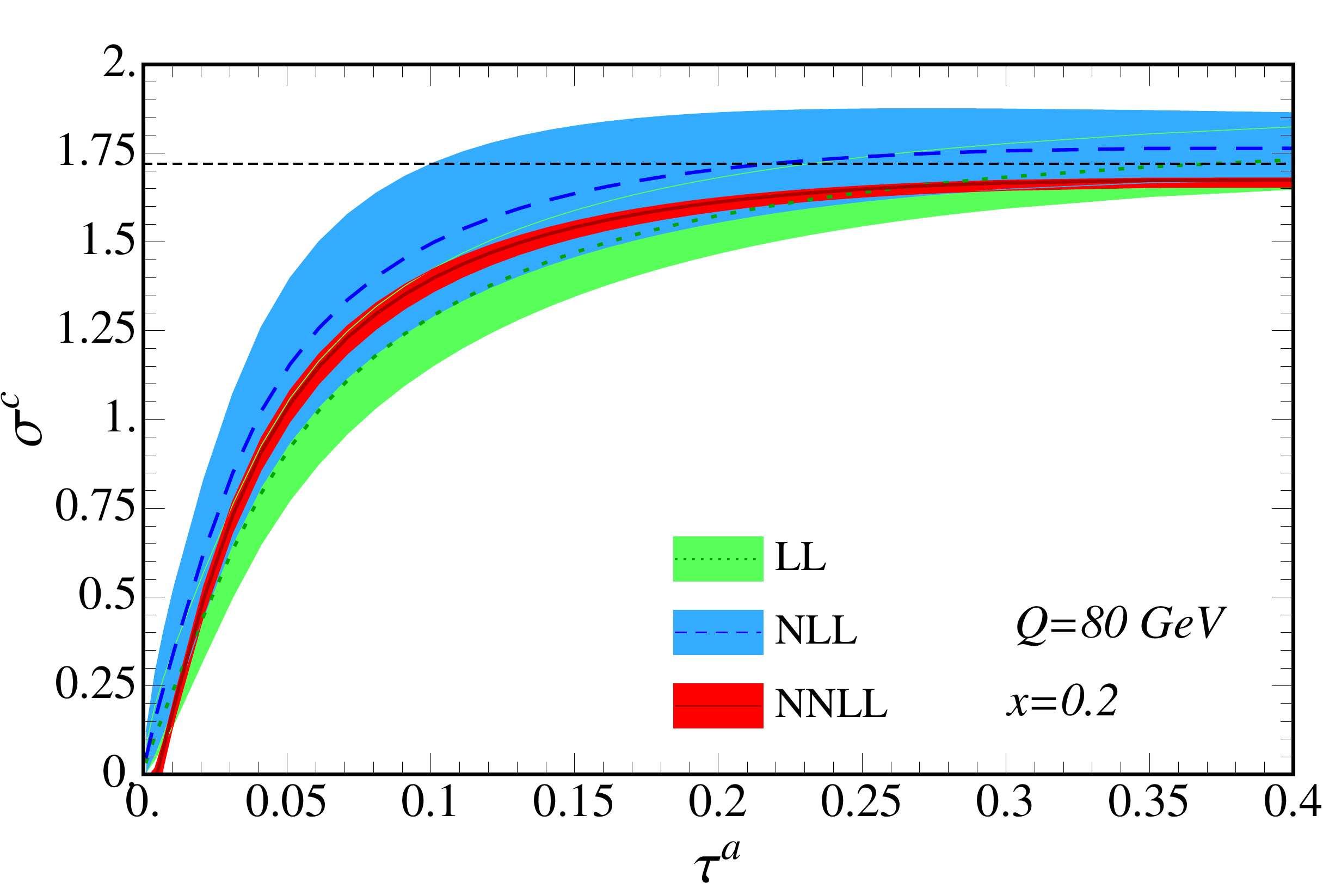}
      \hspace{1em}
      \includegraphics[width=0.45\columnwidth]{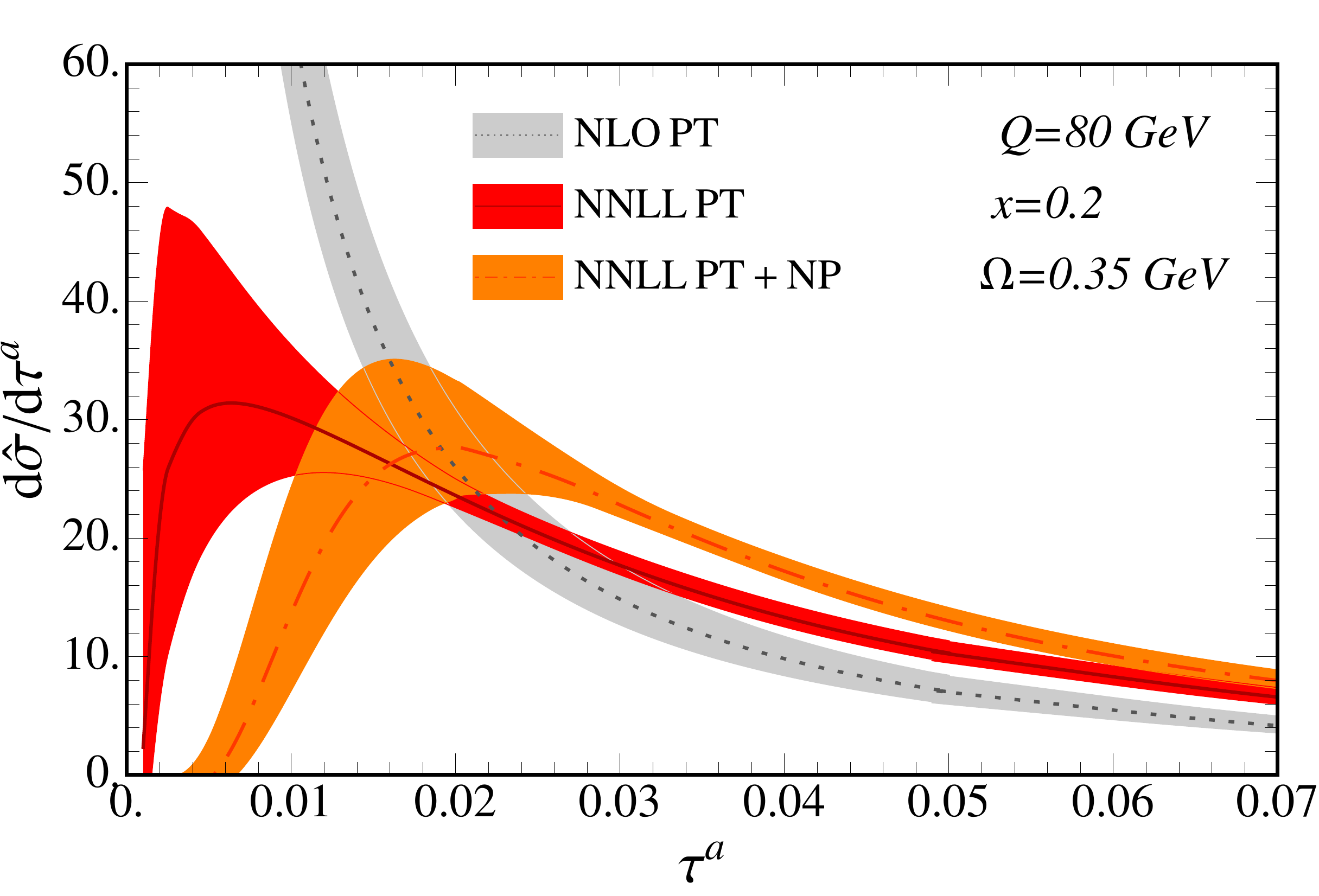}
                      { \caption[1]{ Left: Cumulant distribution in
                          $\taum$. Colored bands and central lines
                          show theoretical uncertainties and central
                          values to LL (dotted line, green band), NLL
                          (dashed line, blue band), and NNLL (solid
                          line, red band) accuracy and the horizontal
                          dashed line is the total cross section.
                          Right: Differential distribution in $\taum$
                          in the peak region, NNLL with
                          nonperturbative shape function taken into
                          account (NNLL PT+NP, dashed, orange), and
                          without NP shape function at fixed-order
                          $\alpha_s$ (NLO PT, dotted, gray) and
                          resummed (NNLL PT, solid, red).}
  \label{fig:taum}
} }
\end{figure}

\begin{figure}[b!]{
      \includegraphics[width=.45\columnwidth]{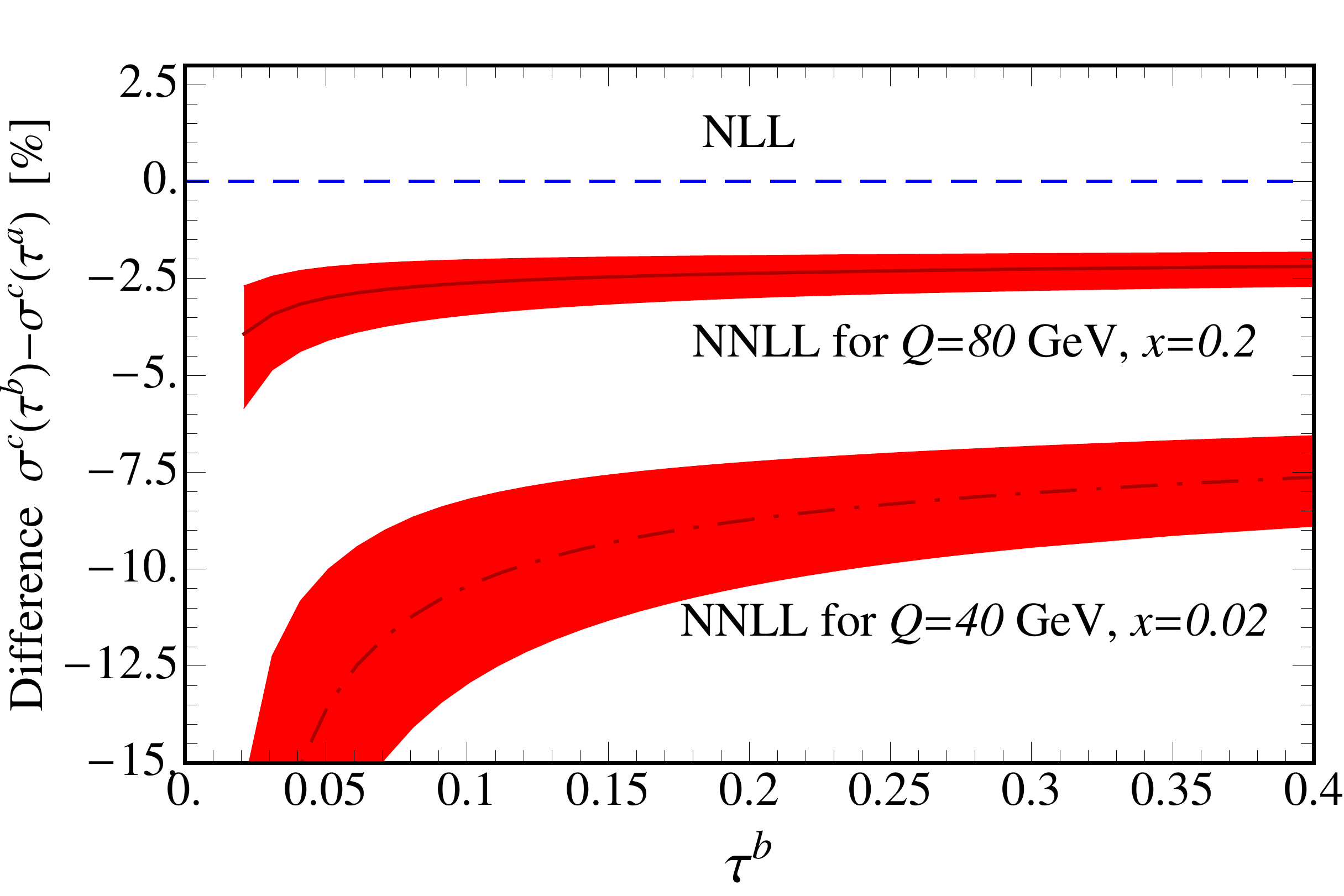}
      \hspace{2em}
      \includegraphics[width=.425\columnwidth]{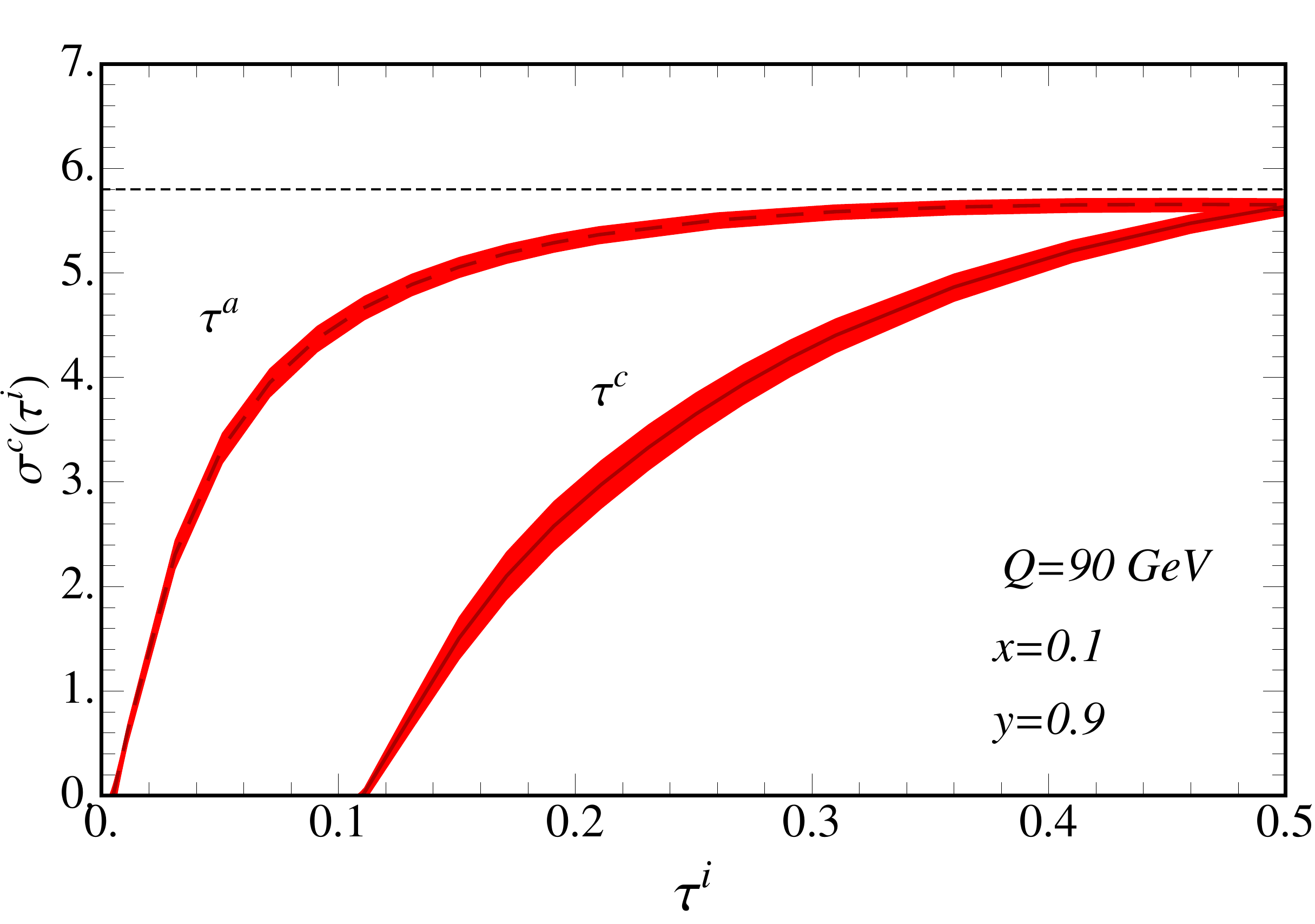}
                      { \caption[1]{ Left: Difference between $\tauB$
                          and $\taum$ cumulant distributions at 2 sets
                          of $Q$ and $x$. The difference vanishes at
                          NLL accuracy.  Right: $\tauCM$ cumulant
                          distribution in comparison to $\taum$
                          distribution.  Notice that $\tauCM$
                          distribution has a threshold at $1-y=0.1$.
                          The horizontal dashed line is the total
                          cross section.  }
  \label{fig:tauBCM}
} }
\end{figure}

In the calculations of \eq{cumulant}, the matrix elements $H$, $J$,
$B$, and, $S$ are evaluated at their natural scales $\mu_{H,B,J,S}$,
at which logarithms in their fixed order calculations are minimized,
and are then evolved to a common scale $\mu$. For example, the natural
scale for $\mu_S$ is $\tau_1 Q$.  The evolution sums up $\log\tau_1$
terms, which is important when $\tau_1\ll1$.  For very small
$\tau_1\sim \Lambda_{\rm QCD}/Q$ the scale $\mu_S$ approaches the
nonperturbative region, and for values in this region must be frozen
at a fixed scale $\sim 1\, {\rm GeV}$ since otherwise the perturbative
expansion for the soft functions anomalous dimension fails. For large
$\tau_1\sim1$ the logs are not large and the resummation must be
turned off so that the fixed order NLO result is reproduced.  The
scales $\mu_{H,B,J,S}$ must change with $\tau_1$ to meet these
requirements, which is achieved with ``profile functions'' which are
used for the results in \fig{taum} and \fig{tauBCM}.  Perturbative
uncertainties are computed by varying all scales up/down by factors of $2$, as
well as by other independent variations of the various scales, and the
uncertainty from these variations decrease as the order in resummed
perturbation theory increases. Explicit expressions of the profile
functions and the variations used are given in
Ref.~\cite{Kang:2013nha}.

\fig{taum} shows the $\taum$ cumulant (Left) and differential (Right)
distribution at $Q=80\GeV$ and $x=0.2$. Three curves represent the
results resummed to LL, NLL, and NNLL accuracy and their perturbative
uncertainty bands.  The plot shows an excellent order-by-order
convergence from LL to NNLL order.  One also finds only a small difference
between the total cross section at ${\cal O}(\alpha_s)$ (dashed
horizontal line) and the NNLL cumulant at large $\taum$, indicating
that the singular terms dominate. (The remaining difference estimates
the size of the small nonsingular terms not taken into account in this
work.)  In the differential distribution, the NNLL result with and
without nonperturbative effects (NNLL PT + NP and NNLL PT) is
presented in comparison with purely fixed-order NLO results (NLO PT).
For the purpose of illustrating the nonperturbative effect, we use the
simplest shape function with a single basis function and
$\Omega_1=0.35\GeV$ for the nonperturbative parameter and convolved
the perturbative cross section with the shape function. The dominant
effect is a shift to the cross section's $\tau_1^a$ value. Above the
peak region this correction reduces to the simple power correction in \eq{powercorrection}
determined by $\Omega_1\sim{\cal O} (\Lqcd/\taun Q)$. For a more
realistic peak region analysis, a shape function with more basis
functions should be used and parameters in the basis should be
determined from experimental data.  In the endpoint region, the NLO
result blows up while the NNLL result is well behaved due to the
resummation of large logs.

The $\tauB$ cross section differs from $\taum$ by a single term at
NLO, which contains $\ln z$. The term is convolved with PDF and
integrated over from $x$ to 1 and its contribution to the cross
section is larger for smaller $x$.  This is shown in the left panel in
\fig{tauBCM}, which displays the percent difference at NNLL for two
sets of $(Q,x)$ values: $(80~{\rm GeV},0.2)$ and $(40~{\rm
  GeV},0.02)$.  For $x=0.2$ the size of the difference is a few
percent, which is small compared to that for smaller $x=0.02$.  The
difference goes up to 10-15\% at $x=0.02$.  This difference is not
sensitive to $Q$, because of the moderate $Q$ dependence in the cross
section. Note that $\taum$ and $\tauB$ do not differ at NLL because
both their NLL logs and LO cross sections are the same.

The 1-jettiness $\tauCM$ measures a jet close to the $z$ axis
(incoming electron direction) in the CM frame and the factorization
theorem in \eq{factorizationintro} is valid for a jet with small
transverse momentum $q_\perp^2=(1-y)Q^2\ll Q^2$.  By the relation
$y=Q^2/xs$ this means that values of $Q$ and $x$ should be chosen to
satisfy $1-y \ll 1$. Here, we use $Q=90\GeV$ and $x=0.1$ which
corresponds to $y=0.9$.  The right panel in \fig{tauBCM} shows
$\tauCM$ in comparison with the $\taum$ cumulant distribution at
NNLL. The most notable feature is the threshold $\theta(\tauCM -1+y)$
which shifts the $\tauCM$ result. This feature is associated with
positivity of the jet mass $M^2_\text{jet}=(\tauCM-1+y) s_J$ at LO.
In addition to the threshold the $\tauCM$ curve increases more gently
than the $\taum$ curve.  This happens because the normalization factor
for the beam axis $q_B$ in $\tauCM$ differs from that in $\taum$ by a
factor of $1/x$.

\section{Summary}

Factorization theorems for two jets in DIS were derived for three
versions of 1-jettiness $\taumBCM$ and numerical results were obtained
up to NNLL order.  The three 1-jettiness' measure particles relative to
3 different axes: jet axis, $z$-axis in the Breit frame and $z$-axis
in CM frame. This leads to different dependence on transverse
momentum.  The factorization theorem is composed of hard, beam, jet,
and soft functions currently known at an order that allows us to achieve NNLL
accuracy. This means that in $\ln\sigma_c$ we resum terms: $\as L^2 \,
(\as L)^k$, $\as L \, (\as L)^k$, and $\as\, (\as L)^k$ where
$L=\log\tau_1$ and $k\ge 0$.  Nonperturbative effects in the
distribution appear as a power correction determined by a nonperturbative
parameter $\Omega_1$ when $\tau_1 Q \gg \Lambda_{QCD}$, which is
universal for each of $\taumBCM$.  Our results contain the dominant
singular terms appearing for small $\tau_1$.  To be accurate for
larger $\tau_1\sim 1$ we need to include non-singular terms which can
be done by matching the fixed order cross section from the
factorization theorem and full QCD. We leave this matching to future
work.

\begin{acknowledgments}
  
  The work of DK and IS is supported by the Office of Nuclear Physics
  of the U.S. Department of Energy under Contract DE-FG02-94ER40818,
  and the work of CL by DOE Contract DE-AC52-06NA25396 and by the LDRD
  office at Los Alamos.

\end{acknowledgments}

\begin{thebibliography}{99}

\bibitem{Farhi:1977sg} 
  E.~Farhi,
  Phys.\ Rev.\ Lett.\  {\bf 39}, 1587 (1977).


\bibitem{GehrmannDeRidder:2007bj} 
  A.~Gehrmann-De Ridder, T.~Gehrmann, E.~W.~N.~Glover and G.~Heinrich,
  Phys.\ Rev.\ Lett.\  {\bf 99}, 132002 (2007)
  [arXiv:0707.1285 [hep-ph]].

\bibitem{GehrmannDeRidder:2007hr} 
  A.~Gehrmann-De Ridder, T.~Gehrmann, E.~W.~N.~Glover and G.~Heinrich,
  JHEP {\bf 0712}, 094 (2007)
  [arXiv:0711.4711 [hep-ph]].

\bibitem{Weinzierl:2008iv} 
  S.~Weinzierl,
  Phys.\ Rev.\ Lett.\  {\bf 101}, 162001 (2008)
  [arXiv:0807.3241 [hep-ph]].

\bibitem{Weinzierl:2009ms} 
  S.~Weinzierl,
  JHEP {\bf 0906}, 041 (2009)
  [arXiv:0904.1077 [hep-ph]].

\bibitem{Becher:2008cf} 
  T.~Becher and M.~D.~Schwartz,
  JHEP {\bf 0807}, 034 (2008)
  [arXiv:0803.0342 [hep-ph]].

\bibitem{Chien:2010kc} 
  Y.~-T.~Chien and M.~D.~Schwartz,
  JHEP {\bf 1008}, 058 (2010)
  [arXiv:1005.1644 [hep-ph]].

\bibitem{Abbate:2010xh} 
  R.~Abbate, M.~Fickinger, A.~H.~Hoang, V.~Mateu and I.~W.~Stewart,
  Phys.\ Rev.\ D {\bf 83}, 074021 (2011)
  [arXiv:1006.3080 [hep-ph]].

\bibitem{Abbate:2012jh} 
  R.~Abbate, M.~Fickinger, A.~H.~Hoang, V.~Mateu and I.~W.~Stewart,
  Phys.\ Rev.\ D {\bf 86}, 094002 (2012)
  [arXiv:1204.5746 [hep-ph]].

\bibitem{Adloff:1997gq} 
  C.~Adloff {\it et al.}  [H1 Collaboration],
  Phys.\ Lett.\ B {\bf 406}, 256 (1997)
  [hep-ex/9706002].

\bibitem{Adloff:1999gn} 
  C.~Adloff {\it et al.}  [H1 Collaboration],
  Eur.\ Phys.\ J.\ C {\bf 14}, 255 (2000)
  [Erratum-ibid.\ C {\bf 18}, 417 (2000)]
  [hep-ex/9912052].

\bibitem{Aktas:2005tz} 
  A.~Aktas {\it et al.}  [H1 Collaboration],
  Eur.\ Phys.\ J.\ C {\bf 46}, 343 (2006)
  [hep-ex/0512014].

\bibitem{Breitweg:1997ug} 
  J.~Breitweg {\it et al.}  [ZEUS Collaboration],
  Phys.\ Lett.\ B {\bf 421}, 368 (1998)
  [hep-ex/9710027].

\bibitem{Chekanov:2002xk} 
  S.~Chekanov {\it et al.}  [ZEUS Collaboration],
  Eur.\ Phys.\ J.\ C {\bf 27}, 531 (2003)
  [hep-ex/0211040].


\bibitem{Chekanov:2006hv} 
  S.~Chekanov {\it et al.}  [ZEUS Collaboration],
  Nucl.\ Phys.\ B {\bf 767}, 1 (2007)
  [hep-ex/0604032].

\bibitem{Antonelli:1999kx} 
  V.~Antonelli, M.~Dasgupta and G.~P.~Salam,
  JHEP {\bf 0002}, 001 (2000)
  [hep-ph/9912488].

\bibitem{Dasgupta:2002dc} 
  M.~Dasgupta and G.~P.~Salam,
  JHEP {\bf 0208}, 032 (2002)
  [hep-ph/0208073].


\bibitem{Dasgupta:2001sh} 
  M.~Dasgupta and G.~P.~Salam,
  Phys.\ Lett.\ B {\bf 512}, 323 (2001)
  [hep-ph/0104277].

\bibitem{Kang:2013nha} 
  D.~Kang, C.~Lee and I.~W.~Stewart,
  arXiv:1303.6952 [hep-ph].

\bibitem{Stewart:2010tn} 
  I.~W.~Stewart, F.~J.~Tackmann and W.~J.~Waalewijn,
  Phys.\ Rev.\ Lett.\  {\bf 105}, 092002 (2010)
  [arXiv:1004.2489 [hep-ph]].

\bibitem{Thaler:2011gf} 
  J.~Thaler and K.~Van Tilburg,
  JHEP {\bf 1202}, 093 (2012)
  [arXiv:1108.2701 [hep-ph]].

\bibitem{Kang:2012zr} 
  Z.~-B.~Kang, S.~Mantry and J.~-W.~Qiu,
  Phys.\ Rev.\ D {\bf 86}, 114011 (2012)
  [arXiv:1204.5469 [hep-ph]].

\bibitem{Kang:2013wca} 
  Z.~-B.~Kang, X.~Liu, S.~Mantry and J.~-W.~Qiu,
  arXiv:1303.3063 [hep-ph].

\bibitem{Stewart:2009yx} 
  I.~W.~Stewart, F.~J.~Tackmann and W.~J.~Waalewijn,
  Phys.\ Rev.\ D {\bf 81}, 094035 (2010)
  [arXiv:0910.0467 [hep-ph]].

\bibitem{Mantry:2009qz} 
  S.~Mantry and F.~Petriello,
  Phys.\ Rev.\ D {\bf 81}, 093007 (2010)
  [arXiv:0911.4135 [hep-ph]].

\bibitem{Jain:2011iu} 
  A.~Jain, M.~Procura and W.~J.~Waalewijn,
  JHEP {\bf 1204}, 132 (2012)
  [arXiv:1110.0839 [hep-ph]].


\bibitem{Salam:2001bd} 
  G.~P.~Salam and D.~Wicke,
  JHEP {\bf 0105}, 061 (2001)
  [hep-ph/0102343].

\bibitem{Lee:2006fn} 
  C.~Lee and G.~F.~Sterman,
  eConf C {\bf 0601121}, A001 (2006)
  [hep-ph/0603066].

\bibitem{Lee:2006nr} 
  C.~Lee and G.~F.~Sterman,
  Phys.\ Rev.\ D {\bf 75}, 014022 (2007)
  [hep-ph/0611061].

\bibitem{Mateu:2012nk} 
  V.~Mateu, I.~W.~Stewart and J.~Thaler,
  Phys.\ Rev.\ D {\bf 87}, 014025 (2013)
  [arXiv:1209.3781 [hep-ph]].

\bibitem{Dokshitzer:1995zt} 
  Y.~L.~Dokshitzer and B.~R.~Webber,
  Phys.\ Lett.\ B {\bf 352}, 451 (1995)
  [hep-ph/9504219].

\bibitem{Akhoury:1995sp} 
  R.~Akhoury and V.~I.~Zakharov,
  Phys.\ Lett.\ B {\bf 357}, 646 (1995)
  [hep-ph/9504248].

\bibitem{Korchemsky:1994is} 
  G.~P.~Korchemsky and G.~F.~Sterman,
  Nucl.\ Phys.\ B {\bf 437}, 415 (1995)
  [hep-ph/9411211].


\end{thebibliography}

\end{document}